# AN EXPERIMENT TO DETECT GRAVITY AT SUB-MM SCALE WITH HIGH-Q MECHANICAL OSCILLATORS

L. HAIBERGER, M. WEINGRAN, H. WENZ, AND S. SCHILLER

*Institut für Experimentalphysik, Heinrich-Heine-Universität Düsseldorf, 40225 Duesseldorf*
www.exphy.uni-duesseldorf.de

Silicon double paddle oscillators are well suited for the detection of weak forces because of their high Q factor (about $10^5$ at room temperature). We describe an experiment aimed at the detection of gravitational forces between masses at sub-mm distance using such an oscillator. Gravitational excitation is produced by a rotating aluminium disk with platinum segments. The force sensitivity of this apparatus is about 10 fN at room temperature for 1000 s averaging time at room temperature. The current limitations to detection of the gravitational force are mentioned.

## 1. Introduction

The existence of a wide energy gap between the electroweak and the Planck scale represents a major problem in modern theoretical physics, for which no solution has been yet found in the frame of the Standard Model. This problem inspired the search for a more fundamental theory, which would provide a common framework for all known interactions. Many theories have been proposed in the past few decades. Most of them imply the existence of new forces and particles [1, 2]. As an example, in the supersymmetric theories light scalars, e.g. moduli, have a mass [1]

$$m \approx M_P \left( \frac{M}{M_P} \right)^k, \qquad (1)$$

where $M_P \sim 10^{19}$ GeV is the Planck mass, $M$ is the mass scale connected to the known elementary particles and $k$ is a small positive integer. Some of these theories propose symmetry breaking at energies in the range 1-10 TeV, which would imply masses of the order of $10^{-4}$-$10^{-2}$ eV for $k = 2$ [3].

Superstring theory (or M-theory), which in principle can describe all foundamental interactions, requires the existence of extra spatial dimensions, that are supposed to be of size comparable to the Planck length ($\sim 10^{-33}$ cm). This length scale is at present not accessible to any experiment.

An alternative approach was proposed a few years ago [3,4,5]. In this model the new spatial dimensions are compactified to the size $r_c$ given by

$$r_c = \frac{1}{M} \left( \frac{M_P}{M} \right)^{2/n}, \qquad (2)$$

where *n* is the number of extra dimensions. If supersymmetry breaking takes place at low energy (~ 1 TeV), the radius $r_c$ can then be much larger than the Planck length as shown in Table 1.

Table 1. Compactification radius as a function of the number of extra dimensions calculated using Eq. (2).

|       | $n = 1$     | $n = 2$    | $n = 3$    |
|-------|-------------|------------|------------|
| $r_c$ | $10^{13}$ m | $10^{-3}$ m | $10^{-9}$ m |

In this model the electro-weak and strong interactions are assumed to be confined to a four-dimensional brane, while gravity turns out to be the only force able to probe the bulk. This assumption implies a modified gravitational potential at distances smaller than $r_c$

$$V = -G_{(4+n)} \frac{m_1 m_2}{r^{1+n}}. \tag{3}$$

Here $G_{(4+n)}$ is the gravitational constant in 4 + n dimensions. Equation (3) can be derived from the Gauss law, when applied to a (4 + n)-dimensional space.

At distances comparable to the compactification radius, Yukawa-type corrections to the Newtonian potential are expected to appear,

$$V = -G \frac{m_1 m_2}{r} \left(1 + \alpha e^{-r/\lambda}\right), \tag{4}$$

where $\alpha$ is a universal constant and $\lambda$ is the range of the Yukawa interaction, which must be of the same order of magnitude as $r_c$. While the scenario with one extra dimension is ruled out by measurements of gravity at planetary distances [6], the case *n* = 2 is not yet completely excluded. A hypothetical deviation from Newton's law could be detected by a laboratory experiment designed to detect gravity in the submillimeter range. In this way it is possible to put a constraint on the unification energy using Eq. (2). The present limits, obtained by laboratory measurements of gravity [2] are shown in Figure 1.

Inspired by these considerations, we have developed an apparatus with the goal of testing gravity in the distance range between 1 and 0.05 mm. In the present work we will describe the principle on which our experiment is based. Then we will describe its major components and their features. Experimental results will be presented and discussed. The last part of this work is devoted to the discussion of possible systematic effects of the present setup and to the proposal of a new experiment, which is designed to have its maximum sensitivity to non-Newtonian effects at distances below 100 μm.

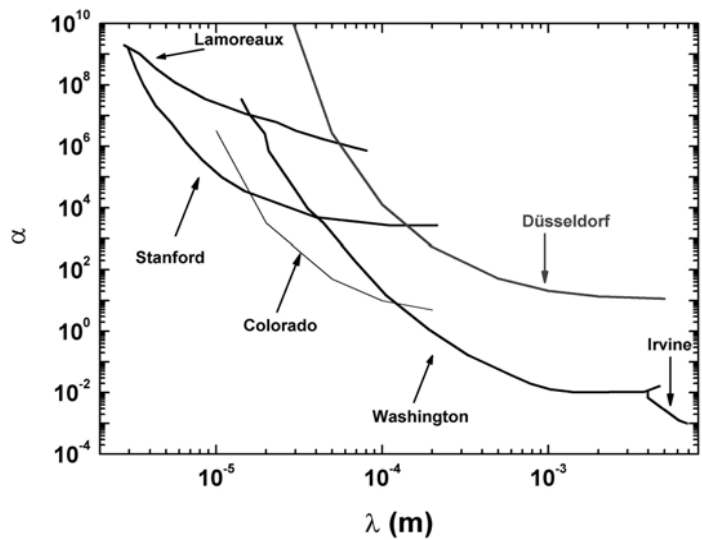

Figure1. Upper limits to the strength of a non-Newtonian force relative to gravity for various recent experiments [ 2] and our current results.

## 2. Principle of measurement

Measuring gravity at short distances requires the use of small test masses (typical size comparable to the gap $r$ between them). Since the gravitational force between spherical masses scales as $r^4$, measuring it proves to be quite challenging. For this reason the choice of the test mass geometry is crucial. A planar geometry is extremely convenient [4, 7]. Our experiment combines the advantages of a flat geometry with the increased sensivity of a resonant detector. The sensor is a harmonic oscillator (resonance frequency $\nu_R$), e.g. a thin plate attached to a spring, while the excitation is provided by a high density plate parallel to the sensor, that is driven at the sensor resonance frequency $\nu_R$. If we assume that the interaction between the test masses is $F \sim 1/r^n$, for a small periodic modulation of the source mass position $r = r_0 + d \cos(2\pi\nu_R t)$ the force can be expanded as

$$F \approx \frac{1}{r_0^n}\left(1 - n\frac{d}{r_0}\cos(2\pi\nu_R t)\right). \tag{5}$$

The amplitude of the oscillator motion at the resonance frequency will be proportional to $nd/r_0^{n+1}$. This implies that the nature of the excitation force can be inferred from a measurement of the $r_0$-dependence of the amplitude response of the sensor. The principal advantage of this modulation technique is that the amplitude response will not be affected by disturbances at frequencies far away from the resonance frequency. The choice of a sensor with resonance frequency in the kHz range represents an advantage, since seismic disturbances appear at much lower frequency.

A fundamental limit to the sensor sensitivity is thermal noise. In analogy to Brownian noise of a free particle, the oscillator is excited by the scattering of its own phonons. This continued excitation can be represented as a spectrally white stochastic force (Langevin force). For a torsional oscillator (as used in the experiment described below), the r.m.s. Langevin thermal torque is given by

$$N_T = \sqrt{\frac{4kT}{\tau}\left(\frac{2\pi m w^2 \nu_R}{3Q}\right)}, \qquad (6)$$

where $\tau$ is the integration time (bandwidth of the measurement $\Delta f = \tau^{-1}$), $2m$ the sensor mass, $Q$ the mechanical quality factor, and $w$ is half the width of the detector area. From Equation (6) it follows that it is convenient to use a sensor with a high Q-factor, which is attained by silicon single-crystal oscillators. The thermal noise torque is of the order of $10^{-17}$ Nm for our oscillators at room temperature and $\tau = 1000$ s. This value corresponds to a minimum detectable force of the order of 10 fN. A further improvement of the sensitivity can be achieved increasing the integration time. Figure 2 shows the goal sensitivity of this experiment compared to previous works.

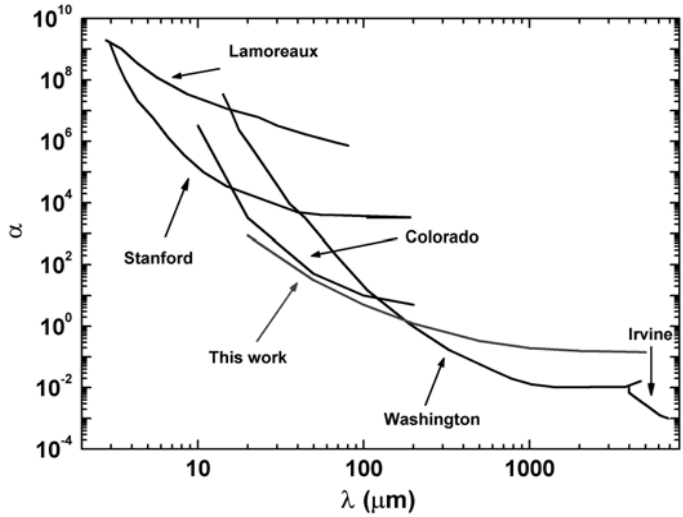

Figure 2. Comparison between the theoretical sensitivity of the experiment described here and the results achieved in previous experiments [ 2]. In the calculations the following parameters were used: $T = 300$ K, $Q = 10^5$, $\tau = 1$ day, $\nu_R = 5$ kHz, $w = 3.65$ mm.

## 3. The sensor

Single-crystal silicon has a very low internal friction and a large knowledge exists concerning its patterning into appropriate geometries. For this reason, we haven chosen this material for the fabrication of our sensors. A sketch of the oscillator used is shown in Figure 3. Its design is similar to the one proposed by Kleiman et al.[8] and improved by Pohl and coworkers [9].
The oscillators were fabricated by us from 300 μm (or 500 μm) thick, float zone refined, double-side polished, <100> oriented, and P-doped silicon wafers with a specific resistance larger than 10 kW·cm. The microfabrication of this sensor was carried out using an alternative technique to the ones used in previous works [18].

The sensor, shown in Figure 3, consists of two paddles, head and wings, connected by a torsion rod, the neck. The wings are themselves connected to a base, foot, by a thinner rod, the leg. The vibrational modes of this structure have been fully characterized in the range between 0.1 and 10 kHz [9]. The one used in this experiment is a torsional mode, known in literature as AS2. Its resonance frequency is ~ 5.5 kHz for 300 μm thick oscillators and about 10 kHz for 500 μm oscillators. In this mode the head rotates twisting the neck, while the wings rotation is out of the oscillator plane around an axis orthogonal to the neck direction [15]. An extremely low internal friction coefficient is typical for the mode AS2 of single–crystal oscillators. This characteristic has already been used to carry out different kinds of precision measurements [9, 12, 13, 15].

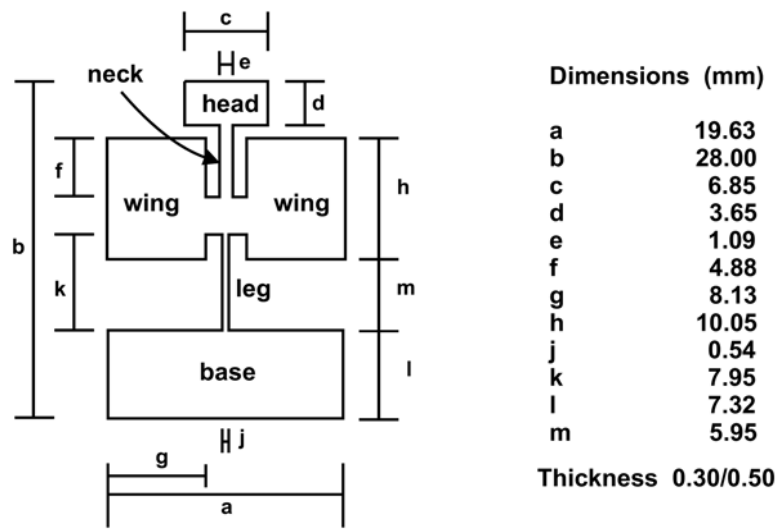

Figure 3. Dimensions of the silicon torsion oscillator.

The detection of the oscillator motion is performed by an optical system, which compared to other detection schemes, provides a very low level of back-action. A laser beam from a He-Ne laser (1 mW optical power) is reflected by the oscillator head and impinges on a position-sensitive photodiode. In the choice of the detection laser it is important to ensure that its power has sufficiently low modulation at the detection frequency, because it is possible to excite the oscillator with light modulated at the resonance frequency [18]. No amplitude modulation is exhibited by the laser used.

The detection system can resolve changes in the displacement amplitude of the oscillator, measured at the resonance frequency, down to $10^{-11}$ m with a bandwidth of 1 Hz. This value corresponds to an angular resolution of $10^{-12}$ rad. As previously shown the amplitude of thermal noise can be reduced by increasing the averaging time $\tau$ (thermal noise scales as $\tau^{-1/2}$). Therefore it is important that the apparatus is stable over long times. Due to thermal and mechanical effects the alignment of the detection beam on the photodiode tends to worsen as the time increases. In order to compensate for this effect the position of the photodiode is actively stabilized with a step motor driven by a proportional controller. This control unit was designed to correct slow drifts, so that they have no influence on the measurement of the oscillator amplitude.

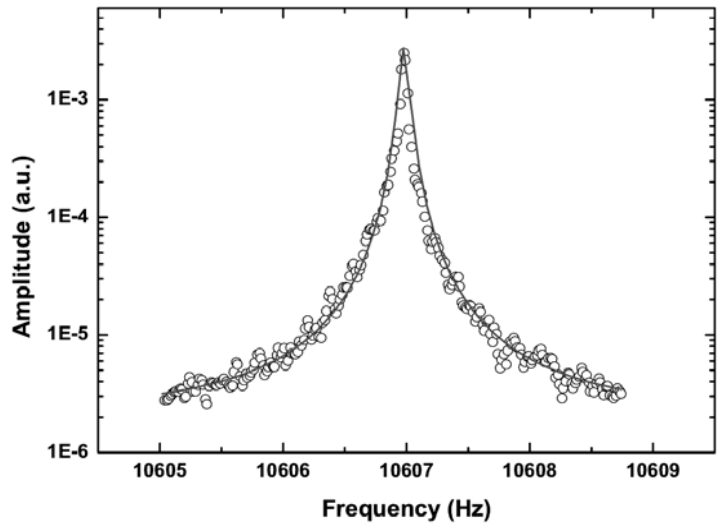

Figure 4. Resonance curve of a 500 μm thick torsional oscillator.

Figure 4 shows an excitation spectrum of a 500 μm thick oscillator taken at room temperature. The resonance curve width is of about 75 mHz. The quality factor, $Q$, can be determined by removing the driving signal and measuring the decay of the amplitude of the oscillations as a function of time, which is given by

$$A(t) = A_0 e^{-\frac{\pi \nu_R t}{Q}}, \qquad (7)$$

where $A_0$ is the stationary amplitude of the oscillator motion when the driving force is applied. The result of a ringdown measurement for a 500 μm thick oscillator is shown in Figure 5. The decay time constant is 22 s, which corresponds to a quality factor of $8 \cdot 10^5$. Typical values of $Q$ for 300 μm thick oscillators are in the range between $1 \cdot 10^5$ and $3 \cdot 10^5$.

In order to achieve an efficient excitation of the sensor it is important that its resonance frequency remains constant over the measurement time. For this reason we studied the influence of temperature drifts on this quantity.

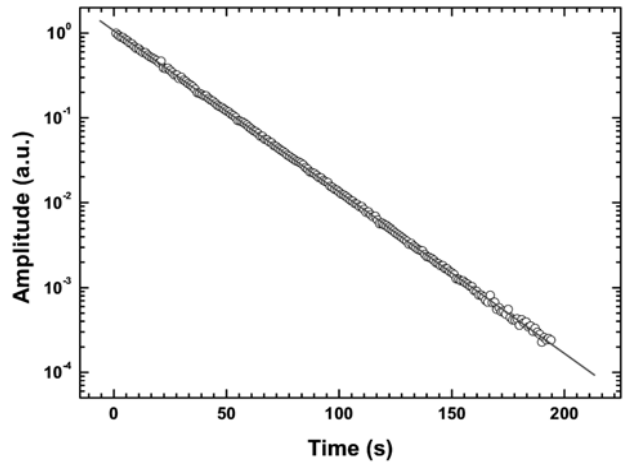

Figure 5. Ringdown measurement for a 500 μm thick oscillator. The resulting decay constant is 22s.

Figure 6 shows the dependence of the resonance frequency of a 300 μm thick oscillator on the temperature. The drift coefficient was 172 mHz/K. Temperature fluctuactions in the laboratory, in spite of the air conditioning system, are on the order of some kelvin. For this reason an active stabilization of the oscillator temperature based on a PID controller was implemented. The obtained stability was better than 0.1 K.

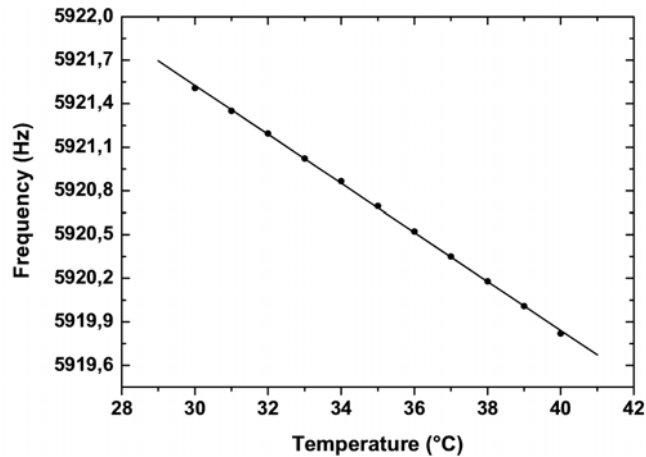

Figure 6. Dependence of the oscillator resonance frequency on the temperature.

## 4. Gravitational excitation

There are two possible ways to produce a gravitational force modulation in front of the sensor. In the first case the excitation is accomplished through position modulation of a source mass along the direction orthogonal to the sensor surface. A disadvantage of this method is that the mass motion can transfer momentum to the oscillator through residual gas molecules excitation (even when the test masses are operated under vacuum condition). A second possibility is to periodically move the source mass along a direction parallel to the sensor.

We followed the second scheme and implemented the source mass motion through a rotating wheel, which holds individual source masses. A sketch of the oscillator/wheel setup is shown in Figure 7.

The wheel is made of a high strength aluminium alloy and is provided with 15 equally spaced holes. These are filled with small platinum disks, that have the same thickness as the wheel. Both wheel surfaces were machined flat, in order to minimize its interaction with residual gas molecules. The platinum masses produce a gravitational attraction at the oscillator resonance frequency when the wheel is rotated with a rotation frequency equal to the $15^{th}$ subharmonic of the oscillator frequency, corresponding to about 20500 rpm. Thus, the vibrations generated by the motor at its rotation frequency and several its harmonics are not resonant with the sensor. The wheel is mounted on a DC brushless motor (Faulhaber). A non-uniform mass distribution in the wheel would reduce the motor lifetime and would cause an increased vibrational noise level, which could couple to the oscillator. For this reason the wheel was balanced. Hereby excess mass was removed from its back, in order to preserve the flatness of the wheel front side.

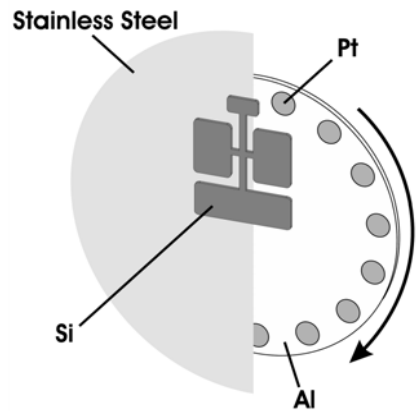

Figure 7. 3D view of the oscillator and source masses attached to the wheel, together with the electrostatic shield. (one half is cut away).

The periodic force excitation due to Newton's gravity was numerically calculated to be of the order of $10^{-14}$ N at a wheel-sensor distance of 100 μm.

To provide an efficient excitation, the wheel rotation frequency must stay constant to within the oscillator linewidth (~35 mHz for a 300μm thick oscillator). This condition requires active stabilization of the rotation frequency. The motor frequency stabilization is achieved via a system consisting of a phase-locked-loop control unit and a commercial servo system. As a reference for the phase-locked-loop we used a function generator with a stability better than 1 ppm in the temperature range 0 to 50 °C. In Figure 8 the relative root Allan-variance of the motor frequency stabilization is shown as a function of the averaging time. The motor has an instability of about $6 \cdot 10^{-8}$ over 400 s, which roughly corresponds to a linewidth ~30 μHz. Since the oscillator linewidth is of 35 mHz, the described motor unit control is well suited for efficient resonant excitation.

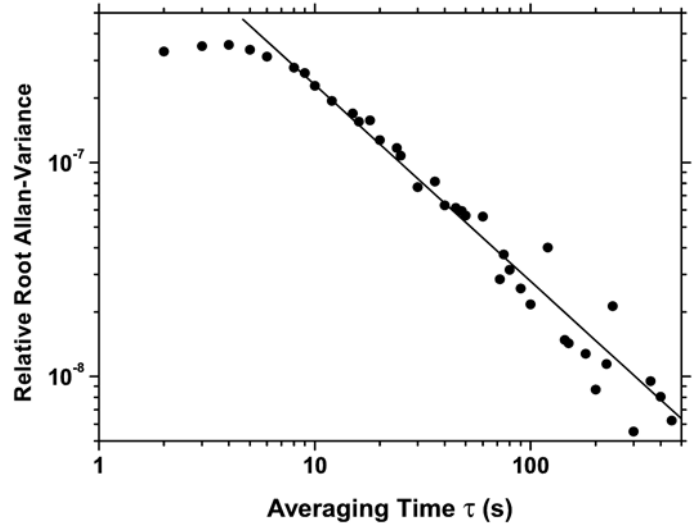

Figure 8. Allan deviation of the frequency-stabilized motor. Data taken over 3600 s.

## 5. Background effects

The most important background effects that can limit the sensitivity of this experiment are vibrations, electrostatic forces, magnetic forces, and pressure-dependent effects due to the presence of residual gas molecules in the vacuum chamber.

The influence of external accelerations in the kHz range can strongly limit the sensitivity of the experiment. The strategy we used to reduce their effects involves several different elements. The vacuum vessels containing the test masses are mounted on top of an optical table. It has a mass of 1.5 tons and is mounted on four pneumatic isolators with a vertical resonance frequency smaller than 1 Hz. Besides seismic accelerations, vibrations generated by the motion of the source mass have to be attenuated. For this reason both motor and oscillator are mounted on separate passive isolation systems, which consist of a stack of three stainless steel disks resting on layers of elastomer springs. The dimensions of the disks have been chosen so that their internal resonances are at frequencies larger than 40 kHz. The elastomer springs are made of RTV 615, a two-component silicon gel, which has been loaded with graphite (~6% graphite by weight) [10, 11]. These dampers show resonances in the range $\nu_0 \sim 10 \div 100$ Hz. The amplitude of an acceleration at frequency $\nu$, well above the damper resonance, is damped by a factor $(\nu/\nu_0)^2$ by each stage. In order to further improve the vibration isolation, the positions of the center of mass of the two indipendent isolation stacks are set at a distance of about 50 cm. Moreover, the external vibrations were reduced by making the metal-on-metal contact surfaces as small as possible.

The elimination of electrostatic effects is achieved by placing a grounded stiff conducting shield between the oscillator and the source mass. This shield consists of a 50 μm thick stainless steel foil, which is clamped in a high vacuum flange. At present no magnetic shield has been implemented.

In principle the wheel motion might couple to the oscillator through residual gas momentum transfer or by some pneumatic effect. In our setup this problem is strongly suppressed by the use of a two-vessel vacuum system. The two chambers are evacuated by two independent pumping systems. To avoid damage to the electrostatic shield, the chambers are connected by an open valve during pump-down. At low pressure the connection valve is closed and there is no possibility for the gas molecule from the motor chamber to interact with the oscillator. Moreover, the thin separation foil between the test masses is sealed with high vacuum sealant and tightly clamped, to reduce its motion.

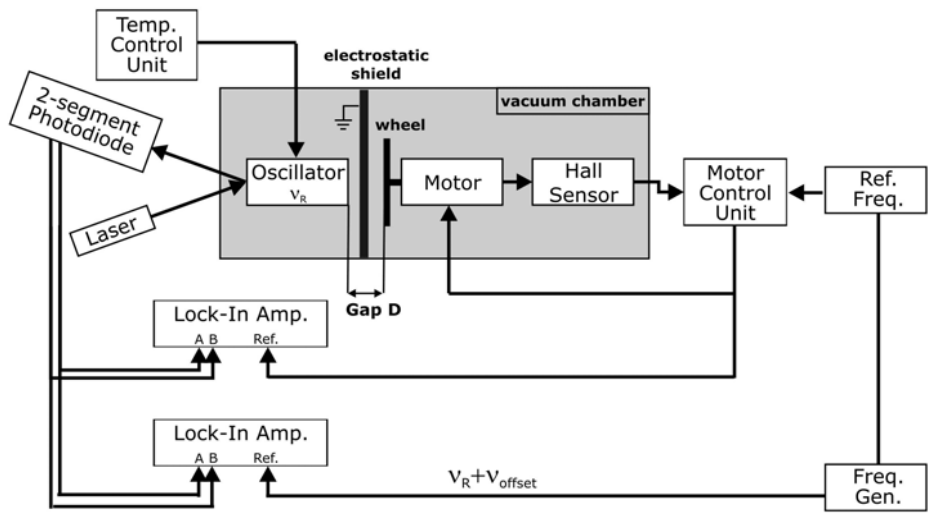

Figure 9. Schematic view of the experiment.

## 6. Experimental results

An overall scheme of the experimental setup is shown in Figure 9. The typical pressure in the oscillator vacuum chamber was of $4 \cdot 10^{-7}$ mbar. The sensor had a resonance frequency of 5921,3 Hz and a quality factor $Q = 1.65 \cdot 10^5$. Both values, measured repeatedly during several runs using different kinds of excitations, were found to constant in time. The oscillator amplitude signal was acquired independently by two lock-in amplifiers. Their reference signals were respectively provided by the motor control unit and by an external function generator, which was set at a different frequency to detect off-resonant noise. Figure 10 displays the phase plots of the oscillator quadratures in absence of any external excitation. The comparison of the two signals statistics shows the contribution of the thermal noise at resonance. The off-resonance noise is due to the detection system.

A typical measurement run followed the following scheme. The oscillator frequency is determined using an external excitation by two small piezoshakers mounted in the oscillator vacuum chamber. The measured frequency value was transfered to the motor frequency stabilization electronics. Next, the gap between sensor and wheel is set. Each of them is mounted in a metal frame provided with four micrometer screws whose tips were adjusted to be in the same plane at a definite distance from the respective mass surface. To set the gap it is sufficient to move the masses until the screw tips touch the electrostatic shield. In this way a common origin for the distance measurements is set. The presence of four different contact signals enables us to check the parallelism of the test masses to the electrostatic shield. Both test masses were then moved back. Since the wheel is moved by a calibrated step motor, its distance values is derived from the step number. The oscillator holder was provided with an eddy current sensor which determines its distance from the metal foil between 0 and 2 mm with a resolution of about 1 μm. The minimum gap between wheel and electrostatic shield is limited by the wheel wobble, while it is rotating. This wobble was measured to be smaller than 10 μm. Next, the quality factor of the oscillator is determined by a ringdown measurement.

Before starting the motor, the oscillator frame is grounded. In order test the distance dependence of the measured signal, the oscillator motion amplitude is measured for different positions of the motor, which was moved stepwise. Each step was of 0.1 mm. The single measurement had a duration of 1200 s. The measured torque as a function of the oscillator-wheel distance is shown in Figure 11. The line represents the expected gravitational torque. It can be seen that the measured signal exceeds the expected gravitational signal of a factor between 20 and 30 depending on the gap. The signal has a functional form similar to the expected one, but it decreases more rapidly with the gap width. This result was reproducible. Rotating the wheel at a different frequency enable us to exclude the existence of off-resonant effects. We also determined that the torque did not depend on the absolute position of the oscillator with respect to the shield.

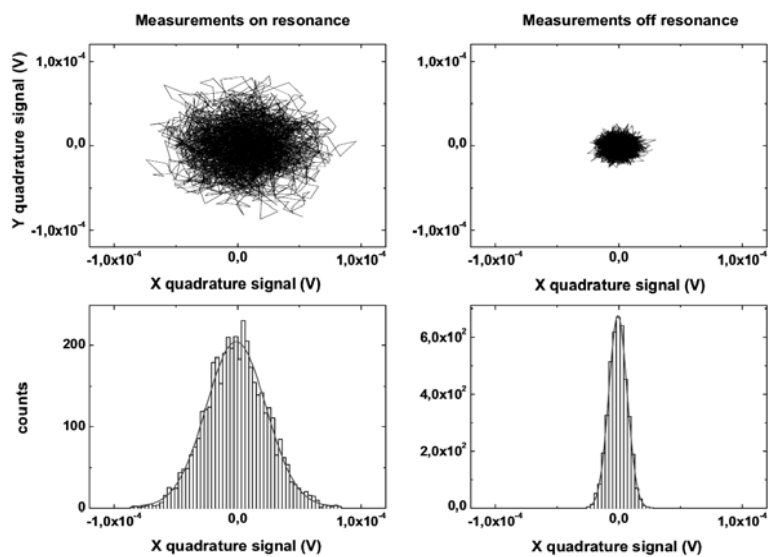

Figure 10. Upper diagrams: phase space trajectory of the oscillator at room temperature driven by thermal noise compared to the trajectory measured at a frequency shifted by 1 Hz (about 30 times the oscillator linewidth) from resonance. Lower diagrams: histograms of the X-quadrature values. Data taken over 4200 s.

To determine the origin of the measured force, several consistency checks were performed. First, a small accelerometer (sensitivity 1 V/g) was mounted on the base plate on which the sensor vibration isolation system stood. No acceleration at the oscillator resonance frequency was detected.

The measured torque did not show any dependence on the pressure in the sensor chamber over the range $10^{-5}$ to $10^{-8}$ mbar. Another possibility considered, was that the movement of the separation foil was excited by the wheel and could then excite the sensor. To test this hypothesis three metal shields with different thicknesses (0.05, 0.1, 0.5 mm) were used. The amplitude of the torque was independent of the shield thickness. A dependence of the quality factor and of the resonance frequency on the distance could also be excluded.

We observed the torque value to be sensitive to the relative position of sensor and source masses. When the oscillator head was higher than the wheel edge, a strong reduction in the torque signal was measured.

A possible electrostatic interaction between the test masses was also investigated. A small electrode (about the size of half an oscillator head) was mounted at the place of the motor. An a.c. voltage at the oscillator resonance frequency was then applied to the electrode. No excitation could be detected for voltages up to 10 V. In order to exclude the possibility that the measured interaction was caused by surface potentials, the wheel was coated with a thin graphite layer (thickness < 1 μm). No change in torque was found.

The electrode was replaced by a small coil to test if an a.c. magnetic field could interact with the sensor. In this case it was possible to detect an excitation of the oscillator. The amplitude of this motion was the same as obtained with the wheel, when the coil generated a field of about 100 mG at the distance of 0.5 mm. The field modulation generated by the motor was found to have an amplitude of about 3 mG at 5921 Hz. While at the motor frequency (about 390 Hz) the amplitude was 3 G. Even if this field, measured at a distance of 0.5 mm from the motor, seems to be to small to be responsible for the oscillator excitation, we cannot exclude that the field measured at the motor rotation frequency may excite the sensor. Investigations in this direction are in progress.

In Figure 1 the present status of this experiment is summarized and compared to the ones of the previuosly cited experiments. From our experimental results the existence of a non-Newtonian correction with $\alpha=20$ and $\lambda=1$ mm can be excluded, so far limited by an unknown systematic effect.

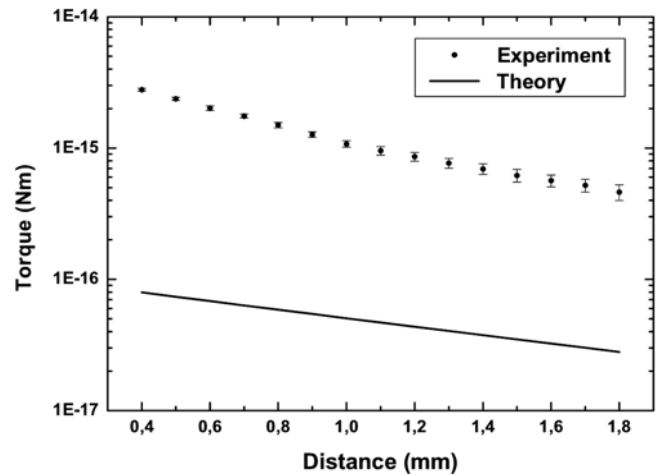

Figure 11. Measured torque compared to the expected gravitational signal (line). Each data point taken over $\tau = 1200$s.

## 7. Proposal for a ultrashort distance experiment

As shown in Figure 1, the existence of extra forces that couple to masses has been poorly constrained in the distance range below 10 μm. The major difficulties in performing an experiment at those distances are due to electrostatic and Casimir force. The impossibility to use a thin metallic plate as shielding, as done previously, requires the development of new concepts. Inspired by a work by Krause and Fischbach [4], we propose a new setup, that could improve the present limits in the range between 0.1 and 100 μm.

A view of the setup is shown in Figure 12. The sensor is a torsional oscillator (labelled 3) similar to the one used in our experiment. The torsion axis is othogonal to the picture plane. The gravitational excitation is provided by two source oscillators of the same type (labeled 1 and 2 in the picture), whose resonance frequencies are equal to $\nu_R$. All three oscillators are parallel. The source oscillators are aligned, while the test oscillator is shifted by one half of its width on one side. The source oscillators' heads are coated with a $t = 1$ μm thick gold layer on the inner side. The test oscillator is coated on one side with a $t = 1$ μm thick gold layer and on the other with 1 μm copper layer. The accesible part of the test oscillator is used to perform the optical detection of its motion. We assume that the gap between each pair of masses has the same size.

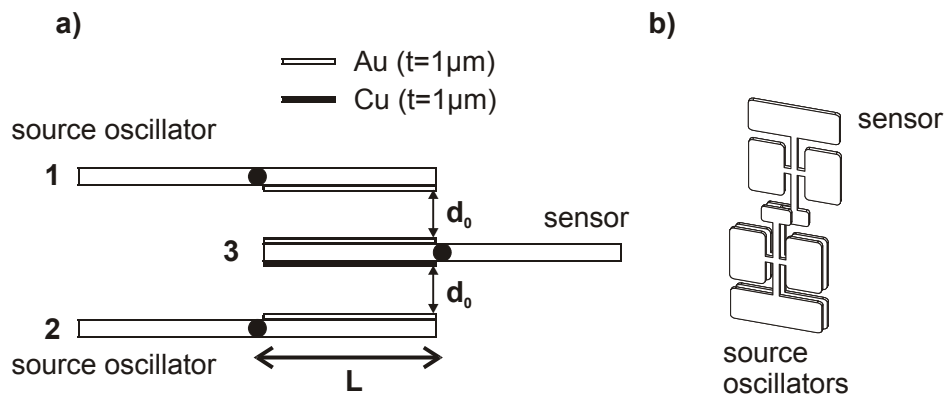

Figure 12. a) Top view of the setup. b) 3D view of the setup.

The source masses are actuated sinusoidally in phase at the frequency $\nu_R$, so that

$$d_1(t) = d_2(t) = d_0 + d_1 \cos(2\pi \nu_R t). \tag{8}$$

This can be accomplished by optical actuaction of the oscillators [18].

The total torque acting on the sensor is then given by

$$\begin{aligned}\tau_{total} &= \tau_{gravitation} + \tau_{electrostatic} + \tau_{casimir} \\ &= \frac{L}{2}\Big[\big(F_1^{gravity} - F_3^{gravity}\big) + \big(F_1^{electrostatic} - F_3^{electrostatic}\big) \\ &\quad + \big(F_1^{casimir} - F_3^{casimir}\big)\Big]\end{aligned} \tag{9}$$

where $F_i^{gravity}$, $F_i^{electrostatic}$, and $F_i^{casimir}$ are the gravitational, electrostatic, and Casimir forces exerted by the oscillators $i = 1$ and $3$ respectively. Under the assuption that $L \gg d$, the gravitational torque, which is due solely to the unequal films on oscillators 1 and 3, can be written in the simple form

$$\tau_{gravitation} = \pi G L^2 H t^2 \rho_{gold} \big(\rho_{gold} - \rho_{copper}\big) \tag{10}$$

Here $G$ is the gravitational constant, the product $L$ times $H$ is the sensor area, $t$ is the thickness of the metal layer, $\rho_{gold}$ and $\rho_{copper}$ are densities of the gold and copper layers. Using the parameters given in Table 2, it results that the gravitational torque $\tau_{gravitation}$ is equal to $2.1 \cdot 10^{-21}$ Nm.

Table 2. Proposed experimental parameters.

| | | |
|---|---|---|
| $t$ | $1 \cdot 10^{-6}$ | m |
| $d_0$ | $2 \cdot 10^{-6}$ | m |
| $d_1$ | $1 \cdot 10^{-6}$ | m |
| $\rho_{gold}$ | $2 \cdot 10^4$ | g/m$^3$ |
| $\rho_{copper}$ | $9 \cdot 10^3$ | g/m$^3$ |

In order to suppress the electrostatic interaction due to charge excess on the metallic layer, a technique can be used that was developed for the spinning test masses of the GP-B satellite and relies on the use of photoemission from UV light [16]. The charges are removed through UV laser light exposition of all the test masses. Moreover in a recent work it was shown that the use of metallic layers realized with plasma-enhanced chemical vapour deposition reduces the effect of surface potentials [17]. Therefore, it should be possible to make $\tau_{electrostatic}$ smaller than the gravitational torque using these two techniques.

If the metal layers were perfect mirrors, the Casimir force on the test oscillator would be perfectly balanced. Metals can be approximated as perfect mirrors at wavelengths larger than their plasma wavelengths, about 0.3 μm for copper and gold [14]. Therefore if the gap between the oscillators are larger than a few times the plasma wavelength, the Casimir contributions of oscillator 1 and 3 should be equal and $\tau_{casimir}$ would vanish. Other background effects, like gravitational attraction between the oscillator bulk (as opposed to the metal layer) and pressure dependent effects, are expected to be perfectly compensated by the symmetry of the proposed setup.

Once all systematic effects are removed, the fundamental limit to the sensitivity of this apparatus is given by thermal noise. For this reason it is convenient to perform the experiment at liquid helium temperature, at which the quality factor of the oscillator reaches values of the order of $10^8$ [19]. The Langevin thermal torque given by Equation (6), of the order of the gravitational torque for an integration time $\tau = 10^6$ s.

Assuming that a Yukawa-type correction to the Newtonian potential exists, it would act on the detector oscillator as a torque

$$\tau_{yukawa} = 2\pi\alpha G\lambda^2 LH \rho_{gold}(\rho_{gold} - \rho_{copper})\left(1 - e^{-t/\lambda}\right)e^{-d/\lambda}\frac{L}{2} \qquad (10)$$

Setting equal to one the ratio between the Yukawa torque and the thermal noise torque, we can express $\alpha$ as a function of the interaction range $\lambda$. The expected constraint on a deviation from Newton's law are plotted in Figure 13, again assuming $\tau = 10^6$ s and $Q = 6 \cdot 10^7$.

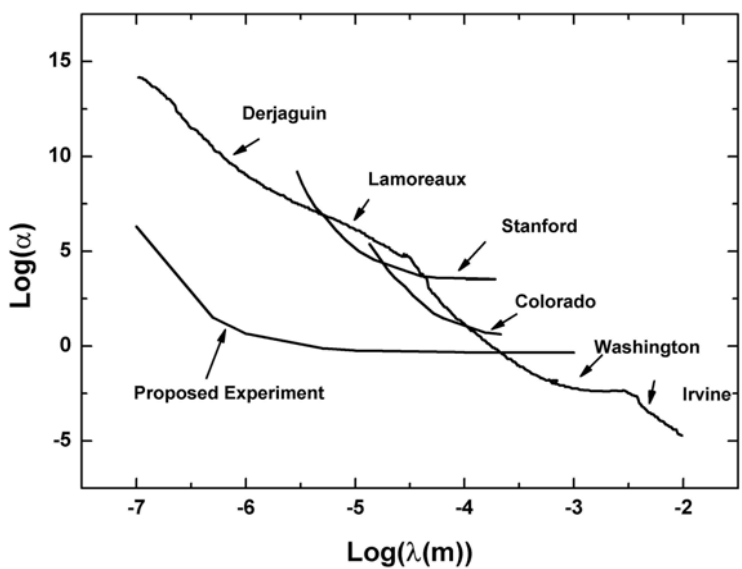

Figure 13. Sensitivity of the proposed experiment compared to present limits.

8. **Conclusions**

The measurement of gravity at laboratory distances can play a role in the search of physics beyond the Standard Model. The use of high-Q mechanical oscillators may help testing the validity of theories predicting the existence of new forces that couple to mass. This is the aim of the experiment presented in this work.

The experimental setup has been characterized. Through stabilization of the sensor temperature and of the excitation frequency this experiment is well suited to long integration times (sensitivity scales as the inverse of the square root of the average time). At present its sensitivity is limited a systematic effect, which has a strenght exceeding of about a factor 30 the expected gravitational signal. Our search for the origin of this systematic effect leads us to believe in the existence of a magnetic coupling between the electric motor and the sensor.

We also proposed the implementation of an experiment suitable for improving the present limits to the existence of an extra force at distance smaller than 100 μm.

**Acknowledgements**

We wish to thank D. Jaeger and R. Pohl for fruitful collaborations on which this work is based. We would like to acknowledge our colleagues in our institute for their support. The project has been financially supported by the Gerhard-Hess Program of the German Science Foundation. One of us (LH) thanks the DAAD (German Academic Exchange Service) for a fellowship.